\documentclass{IEEEtaes}

\usepackage{amsmath,amssymb,amsfonts}
\usepackage{array}
\usepackage{booktabs}
\usepackage{cite}
\usepackage{graphicx}
\usepackage{url}
\usepackage{xcolor}
\usepackage{subcaption}
\usepackage{dblfloatfix}
\usepackage{multirow}

\newsavebox{\cprowA}
\newsavebox{\cprowB}

\newcommand{\E}{\mathbb{E}}

\begin{document}

\title{Imaging-Communication Trade-off in VLEO ISAC-SAR Using CP-OFDM}

\author{In-Hyeok Lee}
\affil{Ulsan National Institute of Science and Technology, Ulsan, Korea} 

\author{Kawon Han}
\member{Member, IEEE}
\affil{Ulsan National Institute of Science and Technology, Ulsan, Korea} 


\authoraddress{In-Hyeok Lee and Kawon Han are with the Ulsan National Institute of Science and Technology (UNIST), Ulsan, South Korea (e-mail: \href{mailto:inhyeokee@unist.ac.kr}{inhyeokee@unist.ac.kr}, \href{mailto:kawon.han@unist.ac.kr}{kawon.han@unist.ac.kr}).}

\markboth{LEE ET AL.}{Imaging-Communication Trade-off in VLEO ISAC-SAR Using CP-OFDM}

\maketitle

\begin{abstract}
This paper investigates the imaging-communication trade-off in very-low-Earth-orbit (VLEO) integrated sensing and communication synthetic aperture radar (ISAC-SAR) using cyclic-prefix orthogonal frequency-division multiplexing (CP-OFDM) as the shared waveform.
We develop a unified analytical framework that jointly accounts for random communication payloads, range-dependent CP deficit across the swath, and slow-time-varying platform Doppler over the synthetic aperture.
The resulting signal model separates the coherently retained component from inter-carrier interference (ICI) and inter-symbol interference (ISI) through common subcarrier-coupling coefficients.
By propagating these effects through receive filtering, range compression, and azimuth focusing, we derive image-domain statistics for matched- and reciprocal-filter receivers.
Based on the derived statistics, an effective noise-equivalent sigma zero (ENESZ) is formulated to characterize data-dependent sidelobes, ICI/ISI, and noise enhancement on a common backscatter-equivalent scale.
The analysis reveals that the CP duration acts as a system-level design parameter bringing scalable trade-off.
Increasing the CP improves coherent retention and suppresses range-dependent interference, but simultaneously reduces coherent processing gain and communication payload throughput.
Accordingly, the sufficient-CP duration does not generally coincide with the imaging-optimal CP duration.
End-to-end simulations for a representative VLEO scenario validate the derived image-domain statistics and demonstrate the resulting trade-off between ENESZ and communication throughput as a function of CP duration.
\end{abstract}


\section{Introduction}
\IEEEPARstart{I}{ntegrated} sensing and communication (ISAC) has emerged as a key paradigm for jointly supporting wireless connectivity and environmental sensing through the shared use of spectrum, hardware, and waveforms \cite{sturm2011waveform,liu2022integrated}. 
Among candidate ISAC waveforms, cyclic-prefix orthogonal frequency-division multiplexing (CP-OFDM) is particularly attractive owing to its widespread adoption in modern communication systems \cite{liu2026clutter,han2026next}. 
Moreover, the data-bearing OFDM signals can be directly exploited for sensing without requiring a dedicated radar waveform, making OFDM a standard-compatible waveform for ISAC systems \cite{wild2021joint}.
Most existing studies on OFDM-based ISAC, however, have focused on terrestrial and relatively short-range scenarios, where sensing is functionally defined as the estimation of a small number of target parameters, such as range, Doppler, and angle \cite{xiao2024joint, mirabella2023deterministic}.

The rapid expansion of low-Earth-orbit (LEO) satellite communications and their envisioned integration into 5G/6G non-terrestrial networks have increased interest in low-altitude orbital platforms for ubiquitous and low-latency connectivity \cite{luo2024leovleo, berthoud2022vleo}.
In particular, very-LEO (VLEO) platforms offer an attractive operating regime for spaceborne ISAC because their reduced slant range can improve radar sensitivity for a given transmit-power budget and alleviate communication path loss.
Moreover, sustained VLEO operation imposes stringent size, weight, and power constraints \cite{crisp2020benefits}, further motivating a unified dual-functional payload.
Together with the demonstrated feasibility of VLEO operation~\cite{jaxa2019slats}, these characteristics motivate the VLEO spaceborne ISAC synthetic aperture radar (SAR) setting considered in this work, where data-bearing CP-OFDM is employed as the shared waveform.

Extending OFDM sensing to spaceborne SAR introduces a fundamentally different operating regime.
This is because SAR imaging requires phase-coherent processing of echoes collected over a long synthetic aperture while the platform travels over a substantial distance. 
As a result, the quality of SAR images using communication signals is governed by the random data payload and the interaction between platform motion and the OFDM time-frequency structure. Despite these distinctive characteristics, the fundamental performance trade-offs of data-bearing CP-OFDM in spaceborne ISAC-SAR have remained largely unexplored.
Conventional OFDM-SAR formulations, however, generally employ predetermined or waveform-designed subcarrier coefficients rather than independently varying communication payload \cite{zhang2014irci,zhang2015ofdm}.

Concretely, applying data-bearing CP-OFDM to spaceborne SAR introduces the following coupled mechanisms that govern the resulting image degradation. 
First, the platform velocity produces a time-varying Doppler shift over the synthetic aperture, thereby violating subcarrier orthogonality and generating inter-carrier interference (ICI) \cite{li2001bounds,hakobyan2018novel}. 
Second, the differential two-way propagation delay across a wide swath can become comparable to or exceed the guard interval implemented as CP \cite{zhang2014irci}. 
A CP deficit causes adjacent OFDM symbols to enter the receiver processing window, producing inter-symbol interference (ISI) together with additional ICI due to incomplete symbol capture \cite{park2004efficient}. 
Third, the random communication payload can produce data-dependent sidelobes after receive filtering, whose level depends on the constellation and the receiver processing \cite{han2026ofdmisac,zheng2024random}.
Together, these mechanisms lower the peak-to-floor ratio of the focused image.

Previous studies have primarily addressed these mechanisms separately. 
Data-induced random sidelobes have been investigated through constellation shaping and optimization, spectrum inversion, and receive filter design \cite{du2024reshaping,du2026probabilistic,mura2025optimized,liu2025uncovering,yang2024constellation,keskin2025fundamental,rodriguez2023supervised,yang2026constellation,liu2025cpofdm}.
However, these studies mainly consider short fast-time intervals or one-dimensional range profiles, thereby failing to characterize how data-dependent interference propagates through coherent SAR processing.
Communication OFDM waveforms have also been exploited for passive SAR imaging \cite{gutierrez2013wimax, kurnia2025integrated}, and joint communication and SAR imaging using modified OFDM waveforms has been experimentally demonstrated \cite{wang2019first}, establishing the feasibility of communication-enabled SAR imaging but without providing an image-domain characterization of data-bearing OFDM impairments.

Doppler-induced ICI has been extensively investigated in OFDM sensing, primarily for target motion \cite{zhang2015ofdm,keskin2021mimo,zhang2020joint}.
In spaceborne SAR, however, platform motion produces a deterministic Doppler sweep over the synthetic aperture for every illuminated scatterer.
Likewise, prior studies on CP deficit have characterized noise-like ISI/ICI from random payloads \cite{xu2025cplength}, degradation beyond the CP-supported delay region \cite{li2025beyond,wang2025coherent}, and ISI-resistant reference-signal designs \cite{tang2024isi}.
These formulations generally parameterize the excess delay by a single scalar parameter, whereas in a spaceborne SAR swath, the differential delay varies continuously with target range, causing scatterers within the same image to experience different degrees of CP deficit.

Consequently, these strands of prior work leave open a unified characterization of how data-payload randomness, platform-induced Doppler ICI, and range-dependent CP-deficit ICI/ISI propagate through a two-dimensional spaceborne SAR imaging chain.
A joint treatment is essential because the CP duration introduces a system-level trade-off between SAR imaging and communication throughput. 
Increasing the CP mitigates delay-induced interference, but it neither eliminates platform-induced Doppler ICI nor avoids the reductions in communication throughput and coherent processing gain caused by a longer OFDM symbol.

To address these gaps, this paper develops an end-to-end analytical framework for data-bearing CP-OFDM ISAC-SAR in a VLEO environment. The main contributions are summarized as follows:

\begin{enumerate}
\item We derive a unified received-signal model for data-bearing OFDM-based ISAC-SAR that jointly accounts for range-dependent CP deficit and platform-induced Doppler. 
The model explicitly separates the coherently retained component, ICI, and ISI through common subcarrier-coupling coefficients.

\item We propagate the received-signal model through the SAR imaging process to derive image-domain statistics for matched- and reciprocal-filter receivers. 
These expressions quantify the contributions of data-dependent sidelobes, ICI/ISI, and thermal noise to the focused-image floor.

\item We formulate an effective noise-equivalent sigma zero (ENESZ) that maps the image-floor contributions from data-dependent sidelobes, Doppler-induced ICI, CP-deficit-induced ICI/ISI, and thermal noise onto a common backscatter-equivalent scale.

\item We characterize the SAR imaging-communication trade-off by accounting for the effects of CP length, platform velocity, and OFDM constellation, and formulate a CP-design criterion that explicitly balances SAR imaging performance and communication throughput.

\end{enumerate}

Simulation results validate the derived image-domain statistics and ENESZ predictions through end-to-end SAR processing. 
The results further demonstrate the effects of CP length, platform velocity, and OFDM constellation, and illustrate the resulting trade-off between SAR imaging performance and communication throughput.

\section{System Model}
\label{sec:model}

\subsection{VLEO ISAC-SAR Geometry and CP-OFDM Signal Model}

\begin{figure}[t]
    \centering
    \includegraphics[width=0.85\columnwidth]
    {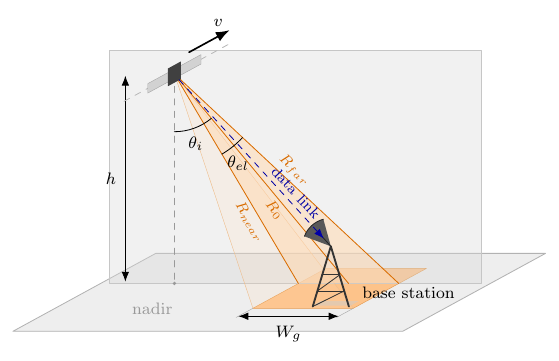}
    \caption{ISAC-SAR geometry in a VLEO orbit.}
    \label{fig:scenario}
\end{figure}

Consider the VLEO ISAC-SAR system, whose imaging-communication geometry is illustrated in Fig.~\ref{fig:scenario}.
The considered system utilizes VLEO communication signals and reuses it for active SAR imaging, allowing a single VLEO platform to provide dual functionality.
The VLEO platform is assumed to support simultaneous transmit-and-receive operation through sufficient transmit-receive isolation and self-interference suppression, as commonly considered in full-duplex ISAC architectures \cite{islam2022fd, le2026fd}.
Residual self-interference is therefore not explicitly modeled in this work.

The platform moves with an effective velocity $v$ at an orbital altitude $h$ and illuminates the ground at an incidence angle $\theta_i$. A zero-squint configuration is assumed, and over the coherent processing interval (CPI), the orbital trajectory is approximated as locally linear. Under a local flat-Earth approximation, the look angle is taken as $\theta_i$, yielding the reference slant range $R_0=h/\cos\theta_i.$
For an elevation beamwidth $\theta_{el}$, the near- and far-range boundaries determine the illuminated slant-range extent
\begin{equation}
\Delta R=
\frac{h}{\cos(\theta_i+\theta_{el}/2)}
-
\frac{h}{\cos(\theta_i-\theta_{el}/2)},
\label{eq:swath}
\end{equation}
with the corresponding ground-swath width $W_g=\Delta R/ \sin\theta_i$.
The differential two-way propagation delay across the swath is therefore
$T_w=2\Delta R / c,$ which defines the echo-window span relevant to the CP requirement.

The transmitted waveform consists of $M$ consecutive CP-OFDM symbols indexed by subcarrier $n\in\{0,\ldots,N-1\}$ and slow-time symbol $m\in\{0,\ldots,M-1\}$. The communication data on the $(n,m)$-th cell is denoted by $d_{n,m}$ and is assumed to be independently drawn from a normalized constellation satisfying
$\mathbb{E}\{|d_{n,m}|^2\}=1.$
The modulated subcarrier symbol is
$s_{n,m}=\sqrt{p_{n,m}}\,d_{n,m}$.
Power is assumed to be uniformly distributed across the $N$ subcarriers within each OFDM symbol, while the symbol-level transmit power $P_m$ is allowed to vary over slow time:
\begin{equation}
p_{n,m}=\frac{P_m}{N}>0,
\qquad
\frac{1}{M}\sum_{m=0}^{M-1}P_m=P_{\rm av},
\end{equation}
where $P_{\rm av}$ denotes the average transmit power.

The subcarrier spacing $\Delta f$ defines the useful symbol duration $T=1/ \Delta f$.
A CP of duration $T_{cp}$ is prepended by cyclically extending the final $T_{cp}$ portion of the useful symbol, yielding $T_s=T+T_{cp}$. The number of symbols $M=\lfloor T_a/T_s \rfloor$ within the CPI decreases as the CP is lengthened, where $T_a$ is the CPI. 
\begin{figure}[t]
    \centering
    \includegraphics[width=0.85\columnwidth]
    {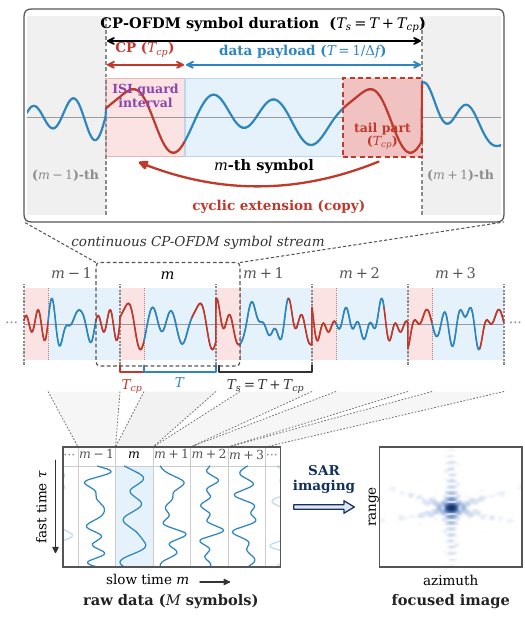}
    \caption{CP-OFDM waveform for ISAC-SAR: the structure of one symbol (top), the continuous symbol stream of period $T_s=T+T_{cp}$ (middle), and the data-payload interval $T$ of each of the $M$ symbols stacked as one column of the raw data and focused SAR image (bottom).}
    \label{fig:transmitted}
\end{figure}
Let $\eta_m$ denote the $m$-th slow time, with $\eta_{m+1}-\eta_m=T_s$. 
The corresponding transmitted baseband waveform is
\begin{equation}
x_m(t)
=
\sum_{n=0}^{N-1}
s_{n,m}
e^{j2\pi n\Delta f(t-\eta_m-T_{cp})},
\label{eq:tx}
\end{equation}
and the complete transmitted waveform is $x(t)=\sum_{m=0}^{M-1}x_m(t)$.
The resulting time-domain structure is in the upper part of Fig.~\ref{fig:transmitted}.

For SAR processing, the observation time is represented in terms of fast and slow time. 
The slow time $\eta_m$ marks the platform position along the synthetic aperture, whereas the fast time $\tau$ is the intra-symbol time that resolves propagation delay in range. 
Each symbol thus acts as one pulse, and the $M$ symbols of the aperture form the columns of the raw data, as shown in the lower part of Fig.~\ref{fig:transmitted}.

\subsection{Swath-Dependent CP Deficit and Platform-Induced Doppler}
\label{sec:model:impairments}

The effects of swath-dependent CP deficit and platform-induced Doppler are investigated in this subsection.
Range gating removes the bulk propagation delay $2R_{\rm near}/c$, such that the CP requirement is governed by the differential swath delay $T_w$ rather than the absolute satellite-to-ground round-trip delay.
Therefore, the differential delay can be defined as
\begin{equation}
\tilde{\tau}_m=\frac{2[R(\eta_m)-R_{\rm near}]}{c},\qquad
0\leq\tilde{\tau}_m\leq T_w.
\label{eq:relative_delay}
\end{equation}
For each symbol, the receiver takes one $T$-duration DFT window, the $m$-th column of the raw data in Fig.~\ref{fig:transmitted}. 
The gate is matched to a reference delay $\tau_g$: the body of a symbol arriving with delay $\tau_g$ begins at $\tau_g+T_{cp}$, so the window spans $\tau\in[\tau_g+T_{cp},\,\tau_g+T_{cp}+T)$. 
A scatterer with $\tau_g\leq\tilde{\tau}_m\leq\tau_g+T_{cp}$ contributes only its own $m$-th symbol to the window, because the CP absorbs the residual offset $\tilde{\tau}_m-\tau_g$. 
We refer to this interval as the ISI-free interval.
The gate placement is not unique and may be selected according to the desired swath coverage.
Throughout this work, we adopt the centered placement $\tau_g=(T_w-T_{cp})/2$, which centers the ISI-free interval
$[\tau_g,\tau_g+T_{cp}]$ within the swath delay span $[0,T_w]$.

For scatterers located outside the ISI-free interval, the desired OFDM symbol is only partially captured by the DFT window. 
The excess delays
\begin{equation}
\delta_+=\max(0,\tilde{\tau}_m-\tau_g-T_{cp}),\quad
\delta_-=\max(0,\tau_g-\tilde{\tau}_m),
\label{eq:excess_delay}
\end{equation}
cannot both be positive. 
A late scatterer ($\delta_+>0$) introduces a portion of symbol $m-1$ at the beginning of the window, whereas an early scatterer ($\delta_->0$) introduces a portion of symbol $m+1$ at its end. 
The desired symbol thus occupies a single contiguous block of length $T-\delta$, with
$\delta=\delta_++\delta_-$, and its fraction of the window, $\beta=1-\delta/T$, is termed the capture factor.
We assume $T_w<T$, which holds for the swath widths and subcarrier spacings considered here.
Therefore, $\delta\leq T_w <T$ and at most one adjacent symbol enters any DFT window.

\begin{figure}[t]
\centering
\includegraphics[width=0.85\columnwidth]
{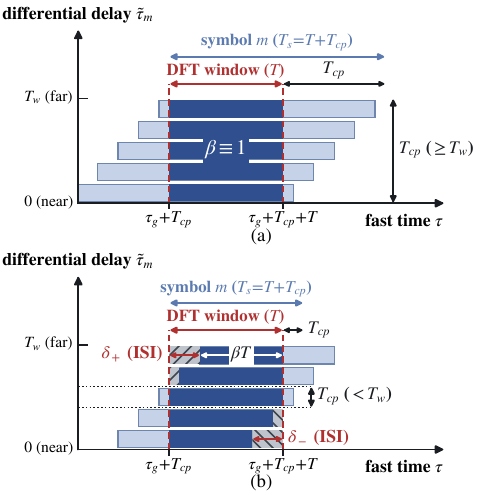}
\caption{Symbol capture scheme at the receive gate for scatterers spanning the swath delay $T_w$: (a) sufficient-CP regime and (b) CP-deficit regime.}
\label{fig:gating}
\end{figure}

The symbol capture mechanism across the swath governed by $T_{cp}$ relative to the swath delay span $T_w$ is illustrated in Fig.~\ref{fig:gating}.
With a sufficient CP, $T_{cp}\ge T_w$, a single gate placement holds every scatterer inside the ISI-free interval, and the entire swath is captured with $\beta=1$ (Fig.~\ref{fig:gating}(a)).
Once $T_{cp}<T_w$, however, some scatterers fall outside the ISI-free interval wherever the gate is placed (Fig.~\ref{fig:gating}(b)).
Under a capture factor $\beta<1$, the desired symbol occupies only the fraction $\beta$ of the window, while the remaining $1-\beta$ is occupied by an adjacent symbol carrying an independent data payload.
A CP deficit therefore substitutes interference for the missing part of the echo rather than merely attenuating it.
The CP duration alone sets the width of the delay interval over which $\beta=1$, but it incurs overhead in every symbol period, thereby linking the swath geometry to both SAR image quality and communication throughput.

Platform motion arises an additional impairment in VLEO ISAC-SAR. 
Over the short duration of an OFDM symbol, the radial velocity is approximated as constant and is given by
$v_r(\eta_m)
=
\frac{dR(\eta)}{d\eta}\bigg|_{\eta=\eta_m}$,
resulting in the Doppler frequency
$f_D(\eta_m)
=
-2v_r(\eta_m)/\lambda.
$
Under the zero-squint geometry considered in this work, the Doppler centroid is zero. Measuring slow time from the zero-Doppler instant and applying a first-order broadside approximation yields
$v_r(\eta)
\simeq
\frac{v^2}{R_0}\eta,
$ and therefore
\begin{equation}
f_D(\eta)
=
-K_a\eta,
\qquad
K_a=\frac{2v^2}{\lambda R_0}.
\label{eq:doppler_history}
\end{equation}
The normalized Doppler offset, which expresses the Doppler shift relative to the OFDM subcarrier spacing, is defined as
\begin{equation}
\varepsilon_m
=
\frac{f_D(\eta_m)}{\Delta f}.
\label{eq:eps}
\end{equation}

\begin{figure}[t]
\centering
\includegraphics[width=0.85\columnwidth]
{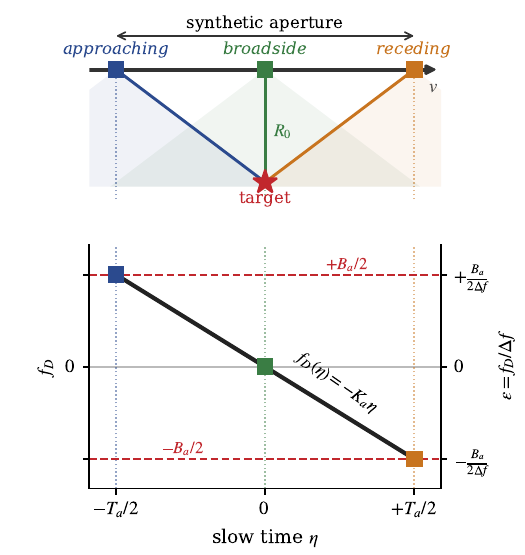}
\caption{Platform-induced Doppler evolution across the synthetic aperture for the scene-center reference scatterer.}
\label{fig:doppler}
\end{figure}

The normalized Doppler offset $\varepsilon_m$ evolves continuously along the synthetic aperture as a consequence of the platform motion, as illustrated in Fig.~\ref{fig:doppler}.
This slow-time-varying offset destroys the exact orthogonality among the OFDM subcarriers and is the platform-induced source of ICI.
Since different scatterers exhibit different range-rate histories and therefore different Doppler shifts, a single common frequency shift cannot simultaneously compensate all echoes.
In contrast, the CP deficit varies across the swath according to the scatterer range and the CP duration.
The two impairments therefore act predominantly along different axes of the raw data: Doppler varies over slow time, whereas the CP deficit varies over range.

\subsection{Received-Signal Model under ISI/ICI Effects}
\label{sec:model:coupling}

In this subsection, a unified received-signal model that incorporates both impairments is introduced in detail.
The receiver discards the CP interval and projects the $m$-th window onto the subcarrier basis over the useful symbol duration $T$, which is the continuous-time form of the DFT taken on that window.
The sample obtained on the $(n,m)$-th cell is
\begin{equation}
y_{n,m}
=
\frac{1}{T}
\int_0^T
r_m(u)
e^{-j2\pi n\Delta f u}
\,du,
\label{eq:demod}
\end{equation}
where $r_m(u)$ denotes the baseband range-gated received waveform during the $m$-th processing interval, and $u=\tau-\tau_g-T_{cp}~ (0\leq u<T)$ is the fast-time coordinate relative to the DFT-window origin.

The channel coefficient associated with the $(n,m)$-th cell of a point scatterer is defined as
\begin{equation}
h_{n,m}
=
\alpha g_m
e^{-j4\pi R(\eta_m)/\lambda}
e^{-j2\pi n\Delta f(\tilde{\tau}_m-\tau_g)},
\label{eq:hnm}
\end{equation}
where $\alpha$ is the complex reflectivity of the scatterer and $g_m$ is a two-way antenna pattern gain at slow time $\eta_m$. For the image-domain analysis, the pattern is taken uniform over the synthetic aperture, $g_m=1$ for the $M$ symbols within the CPI, so that $|h_{n,m}|=|\alpha|$ for every cell of the scatterer.
For notational simplicity, the derivation below is presented for a single point scatterer.
The Doppler phase common to all subcarriers within the DFT window is
$\theta_{D,m}
=
2\pi f_D(\eta_m)(\tau_g+T_{cp}).$
The desired $m$-th symbol occupies $
\mathcal{B}
=
[\delta_+,\,T-\delta_-],$ 
whose duration is $\beta T$, whereas its complement
$\mathcal{C}
=
[0,T)\setminus\mathcal{B}
$
is occupied by the adjacent OFDM symbol when $\beta<1$.

The coupling produced by the desired and adjacent symbols is then described by
\begin{equation}
\tilde I_k(\varepsilon_m)
=
\frac{1}{T}
\int_{\mathcal{B}}
e^{j2\pi(k+\varepsilon_m)\Delta f u}
\,du,
\label{eq:Idef}
\end{equation}
\begin{equation}
J_k(\varepsilon_m)
=
\frac{1}{T}
\int_{\mathcal{C}}
e^{j2\pi(k+\varepsilon_m)\Delta f u}
\,du,
\label{eq:Jdef}
\end{equation}
where $n'$ is the subcarrier on which the symbol was transmitted and $k=n'-n$ the offset from the observed subcarrier $n$.
Since $\Delta fT=1$, the desired-symbol coefficient has the closed form
\begin{equation}
\tilde I_k(\varepsilon_m)
=
\beta
e^{j\pi(k+\varepsilon_m)
\left(1+\frac{\delta_+-\delta_-}{T}\right)}
\operatorname{sinc}
\left[\beta(k+\varepsilon_m)\right].
\label{eq:Ik}
\end{equation}
For the adjacent symbol,
\begin{equation}
J_k(\varepsilon_m)=(1-\beta)e^{j\pi(k+\varepsilon_m)\varkappa}
\operatorname{sinc}[(1-\beta)(k+\varepsilon_m)],
\label{eq:Jk}
\end{equation}
where
\begin{equation}
\varkappa=
\begin{cases}
1-\beta, & \delta_+>0,\\
1+\beta, & \delta_->0.
\end{cases}
\end{equation}
Of particular importance is the coefficient of the intended cell, for which $n'=n$, obtained by setting $k=0$ in \eqref{eq:Ik},
\begin{equation}
\tilde I_0(\varepsilon_m,\beta)
=
\beta
e^{j\pi\varepsilon_m
\left(1+\frac{\delta_+-\delta_-}{T}\right)}
\operatorname{sinc}(\beta\varepsilon_m),
\label{eq:I0}
\end{equation}
whose squared magnitude is
$|\tilde I_0(\varepsilon_m,\beta)|^2
=\beta^2\operatorname{sinc}^2(\beta\varepsilon_m)$.
This coefficient represents the coherent fraction retained by the intended cell under the joint action of CP deficit and platform-induced Doppler.
The remaining energy is redistributed through $\tilde I_{k\neq0}$ and $J_k$ into the ICI and ISI components, respectively.

\begin{figure}[t]
\centering
\includegraphics[width=0.85\columnwidth]{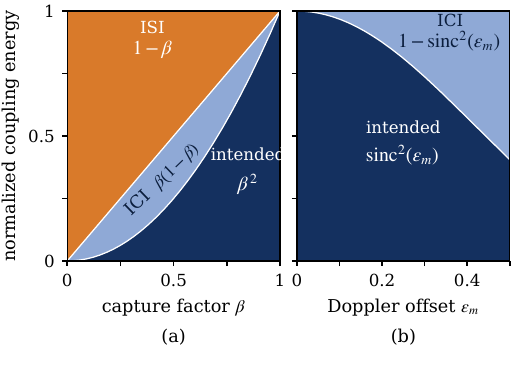}
\caption{Normalized coupling-energy partition in two limiting cases:
(a) CP deficit only ($\varepsilon_m=0$), as a function of the capture factor
$\beta$. (b) Doppler only ($\beta=1$), as a function of the normalized
Doppler offset $\varepsilon_m$.}
\label{fig:partition}
\end{figure}

The coefficients in \eqref{eq:Idef} and \eqref{eq:Jdef} are the Fourier
coefficients on $[0,T)$ of the indicator functions of $\mathcal{B}$ and
$\mathcal{C}$ multiplied by the unit-magnitude factor
$e^{j2\pi\varepsilon_m\Delta f u}$. Considering the complete Fourier basis,
Parseval's relation gives
\begin{equation}
\sum_{k\in\mathbb{Z}}
\big|\tilde I_k(\varepsilon_m)\big|^2
=
\beta,
\qquad
\sum_{k\in\mathbb{Z}}
\big|J_k(\varepsilon_m)\big|^2
=
1-\beta,
\label{eq:energy_partition}
\end{equation}
for every $\varepsilon_m$. Accordingly, the intended cell retains the
normalized coupling energy
$|\tilde I_0(\varepsilon_m,\beta)|^2$, while the coupling energies associated
with ICI and ISI are
$\beta-|\tilde I_0(\varepsilon_m,\beta)|^2$ and $1-\beta$, respectively.
Hence, the normalized coupling energy is partitioned as
\begin{equation}
|\tilde I_0(\varepsilon_m,\beta)|^2
+
\big[
\beta-|\tilde I_0(\varepsilon_m,\beta)|^2
\big]
+
(1-\beta)
=
1.
\label{eq:energy_conservation}
\end{equation}

Accounting for the deterministic phase offset of the adjacent OFDM symbol,
$h^{\pm}_{n,m}
=
h_{n,m}
e^{\pm j2\pi n\Delta fT_{cp}},
$
the $y_{n,m}$ in \eqref{eq:demod} becomes
\begin{equation}
\begin{split}
y_{n,m}
=
&e^{j\theta_{D,m}}
\Bigg[
\tilde I_0(\varepsilon_m,\beta)
h_{n,m}s_{n,m}\\
&+
\underbrace{
\sum_{k\neq0}
\tilde I_k(\varepsilon_m)
h_{n+k,m}s_{n+k,m}
}_{\text{ICI}}\\
&+
\underbrace{
\sum_k
J_k(\varepsilon_m)
h^{\pm}_{n+k,m}
s_{n+k,m\mp1}
}_{\text{ISI}}
\Bigg]
+
z_{n,m},
\end{split}
\label{eq:percell}
\end{equation}
where $z_{n,m}$ denotes the additive thermal noise with variance $\sigma ^2$.
Equation~\eqref{eq:percell} completes the received-signal model of the CP-OFDM ISAC-SAR system. 
In a single expression, it separates the coherently retained component from the ICI and the ISI generated by the CP deficit and the platform-induced Doppler. 
Both mechanisms act through the common coupling kernels $\tilde I_k$ and $J_k$, where the range-dependent $\beta$ and the
slow-time-dependent $\varepsilon_m$ enter jointly in the arguments $\beta(k+\varepsilon_m)$ and $(1-\beta)(k+\varepsilon_m)$. 
For a distributed scene with multiple scatterers, the received cell superposes \eqref{eq:percell} over all illuminated scatterers, each with its own range-dependent $\beta$ and channel coefficient $h_{n,m}$.

\begin{figure}[t]
\centering
\includegraphics[width=0.85\columnwidth]{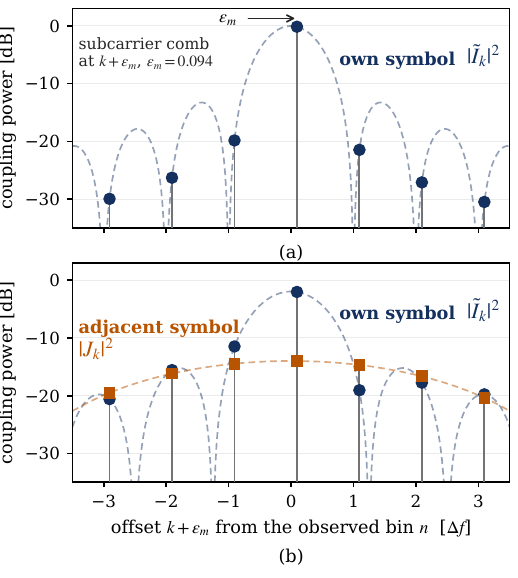}
\caption{Coupling coefficients \eqref{eq:Ik}-\eqref{eq:Jk} as samples of the window responses taken by the transmitted subcarrier comb at offsets $k+\varepsilon_m$: (a) full capture, $\beta=1$. (b) partial capture, $\beta=0.8$.}
\label{fig:coupling}
\end{figure}

The coupling-energy partition in \eqref{eq:energy_conservation} provides a direct interpretation of how CP deficit and Doppler redistribute the received energy.
Under CP deficit alone, the intended, ICI, and ISI contributions are $\beta^2$, $\beta(1-\beta)$, and $1-\beta$, respectively.
Under Doppler alone, the intended coupling energy is $\mathrm{sinc}^2(\varepsilon_m)$, and the remaining $1-\mathrm{sinc}^2(\varepsilon_m)$ becomes ICI.
These two limiting cases are summarized in Fig.~\ref{fig:partition}.
Thus, Doppler redistributes coupling energy from the intended cell into ICI only, whereas a CP deficit redistributes it into both ICI and ISI.

The coupling coefficients can also be interpreted in the frequency domain through the window responses shown in Fig.~\ref{fig:coupling}.
Dashed curves are the responses of the two windows, and the stems are the subcarriers seen from the observed subcarrier $n$. Circles and squares mark $|\tilde{I}_k|^2$ and $|J_k|^2$, respectively.
For full symbol capture, the own-symbol response has zeros at the integer subcarrier offsets, so that orthogonality is recovered when $\varepsilon_m=0$ and $\beta=1$. 
A nonzero Doppler offset shifts the subcarrier sampling locations away from these zeros, thereby producing ICI.
Under partial capture, the own-symbol response is broadened and its zeros no longer coincide with the subcarrier grid, generating ICI even without Doppler. 
At the same time, the truncated adjacent-symbol interval produces the additional $J_k$ responsible for ISI.

\section{CP-OFDM SAR Imaging with Random Communication Data}
\label{sec:sar_imaging}

\subsection{Receive Filtering and Range Compression}
\label{sec:rangecomp}

\begin{table}[t]
  \centering
  \caption{Second-order statistics of the filtered data coefficients
           associated with each term of \eqref{eq:percell}.}
  \label{tab:moments}
  \begin{tabular}{@{}c cc c c c@{}}
    \toprule
    \multirow{2}{*}{\shortstack{Receive\\filter}}
      & \multicolumn{2}{c}{Intended cell}
      & ICI & ISI & Noise \\
    \cmidrule(lr){2-3} \cmidrule(lr){4-4} \cmidrule(lr){5-5} \cmidrule(l){6-6}
      & $\mathbb{E}[\chi_n]$ & $\mathrm{Var}[\chi_n]$
      & $\mathbb{E}[|\xi_{n,k}|^2]$
      & $\mathbb{E}[|\xi^{\pm}_{n,k}|^2]$
      & $\mathbb{E}[|\zeta_n|^2]$ \\
    \midrule
    MF & $1$ & $\mu_4 - 1$ & $1$
       & $\dfrac{P_{m\mp1}}{P_m}$
       & $\dfrac{\sigma^2}{p_{n,m}}$ \\[1.2ex]
    RF & $1$ & $0$ & $\nu$
       & $\nu\,\dfrac{P_{m\mp1}}{P_m}$
       & $\dfrac{\nu\sigma^2}{p_{n,m}}$ \\
    \bottomrule
  \end{tabular}
\end{table}

Since the transmitted symbols $s_{n,m}$ randomly vary across both subcarrier and slow-time indices, the sensing receiver first compensates for the known transmitted communication data through receive filtering, which forms the filtered grid
\begin{equation}
\hat H_{n,m}=\rho_{n,m}y_{n,m}.
\label{eq:filt}
\end{equation}
Throughout the analysis, we investigate two representative frequency-domain receive filters: the matched filter (MF) takes $\rho_{n,m}=s_{n,m}^{*}/p_{n,m}$, and the reciprocal filter (RF) takes $\rho_{n,m}=1/s_{n,m}$~\cite{sturm2011waveform}. 

The constellation is further assumed to be four-fold rotationally symmetric, so that $\mathbb{E}[d^2]=0$. 
The constellation then enters only through $\mu_4=\mathbb{E}[|d|^4]$ and $\nu=\mathbb{E}[|d|^{-2}]$. 
The second-order statistics of the filtered coefficients of the intended cell $(\chi_n=\rho_{n,m}s_{n,m})$, the ICI $(\xi_{n,k}=\rho_{n,m}s_{n+k,m})$, the ISI $(\xi^{\pm}_{n,k}=\rho_{n,m}s_{n+k,m\mp1})$, and the noise $(\zeta_n=\rho_{n,m}z_{n,m})$ are listed in Table~\ref{tab:moments}. 
The ICI carries the transmit power $P_m$ of the desired cell and the power cancels through filtering,
whereas the ISI originates from a different OFDM symbol and retains $P_{m\mp1}/P_m$.

Range compression is conducted by an inverse DFT along the subcarrier dimension,
\begin{equation}
\hat h_m[l]
=
\frac{1}{\sqrt{N}}
\sum_{n=0}^{N-1}
\hat H_{n,m}
e^{j2\pi nl/N}.
\label{eq:profile}
\end{equation}
A scatterer with relative delay $\tilde\tau_m$ responds coherently at the range bin $l_{q,m}=B(\tilde\tau_m-\tau_g)$, taken on the discrete range grid. 

The intended component of~\eqref{eq:percell} remains coherent after range compression. 
Under the assumed independence and four-fold rotational symmetry of the data symbols, the filtering fluctuation, ICI, and ISI terms are zero mean, and their cross terms vanish in the second-order statistics. 
Consequently, they contribute an incoherent range-profile floor rather than coherent replicas of the scatterer.

The mean range-profile power is given by
\begin{equation}
\mathbb{E}\!\left[|\hat h_m[l]|^2\right]
=
N|\alpha_m|^2
|\tilde I_0(\varepsilon_m,\beta)|^2
\delta[l-l_{q,m}]
+
F_m,
\label{eq:range_power}
\end{equation}
where $\alpha_m=\alpha g_m e^{-j4\pi R(\eta_m)/\lambda}e^{j\theta_{D,m}}$, and $F_m$ collects the data-dependent sidelobes, the ICI and ISI leakage, and the thermal noise. 
From Table~\ref{tab:moments}, the per-symbol floors can be written in closed-form expressions as
\begin{equation}
    \begin{split}
        F_m^{\mathrm{MF}}
=
|\alpha_m|^2
\bigg[
|\tilde I_0|^2(\mu_4&-1)
+\beta-|\tilde I_0|^2\\
&+(1-\beta)\frac{P_{m\mp1}}{P_m}
\bigg]
+\frac{\sigma^2}{p_{n,m}},
    \end{split}
\label{eq:floormf}
\end{equation}
\begin{equation}
    \begin{split}
        F_m^{\mathrm{RF}}
=
\nu
\bigg[
|\alpha_m|^2
\big(
&\beta-|\tilde I_0|^2\\
&+(1-\beta)\frac{P_{m\mp1}}{P_m}
\big)
+\frac{\sigma^2}{p_{n,m}}
\bigg].
    \end{split}
\label{eq:floorrf}
\end{equation}

As shown in \eqref{eq:floormf}-\eqref{eq:floorrf}, the MF carries a data-dependent sidelobe proportional to $\mu_4-1$, which the RF eliminates at the cost of scaling the ICI, ISI and thermal noise contributions by $\nu$.
The receiver choice therefore reflects a trade-off between suppressing the data-dependent pedestal sidelobe and preserving robustness to the interference and noise.

\subsection{Azimuth Focusing and Image Floor-to-Peak Ratio}
\label{sec:azimuth}

Backprojection integrates the range-compressed profiles over the synthetic aperture,
\begin{equation}
I(\mathbf r)
=
\sum_{m=0}^{M-1}
\hat h_m\!\left[l_{\mathbf r}(\eta_m)\right]
e^{j4\pi R(\mathbf r,\eta_m)/\lambda},
\label{eq:bpa}
\end{equation}
where $R(\mathbf{r},\eta_m)$ is the slant range to the hypothesized pixel and $l_{\mathbf{r}}(\eta_m)=B[2(R(\mathbf{r},\eta_m)-R_{\rm near})/c-\tau_g]$ is its range bin. 
The propagation phase in~\eqref{eq:bpa} is deterministic, so the target response accumulates coherently across slow time. 
The residual phases $\theta_{D,m}$ and $\arg\tilde I_0$ are approximately linear in $\eta_m$ because $f_D(\eta)=-K_a\eta$. 
A linear phase along slow-time is a Doppler-centroid offset, so it registers the response at a shifted azimuth position and leaves the chirp rate, and hence the focusing, unchanged. 
The offset common to the swath is absorbed into the image grid, and the part varying with $\tilde\tau_m$ is bounded by $vT_w$ in azimuth, well inside one resolution cell $\rho_a$.
In contrast, the pedestal, ICI, ISI, and thermal noise stay mutually incoherent across independently modulated OFDM symbols.
The peak therefore grows as $M^2$ and the floor as $M$, so aperture integration reduces the image floor-to-peak ratio by a factor of $M$.

The aperture-averaged coherent-retention factor is defined as
\begin{equation}
A_0
=
\left\langle
|\tilde I_0(\varepsilon_m,\beta)|^2
\right\rangle_m .
\label{eq:A0}
\end{equation}
For the operating regime considered here, the variation of $|\tilde{I}_0|$ over the synthetic aperture is sufficiently small that $\left\langle |\tilde{I}_0| \right\rangle^2_m\simeq \left\langle |\tilde{I}_0|^2 \right\rangle_m=A_0$.
The normalized interference excess is defined as
\begin{equation}
\bar{\Delta}_P
=
\frac{
\left\langle
\beta-|\tilde I_0(\varepsilon_m,\beta)|^2
+
(1-\beta)\dfrac{P_{m\mp1}}{P_m}
\right\rangle_m
}{
A_0
}.
\label{eq:delta_power}
\end{equation}
Equivalently, $\bar{\Delta}_P$ can be decomposed as
\begin{equation}
\bar{\Delta}_P
=
\left(\frac{1}{A_0}-1\right)
+
\frac{
\left\langle
(1-\beta)
\left(
\frac{P_{m\mp1}}{P_m}-1
\right)
\right\rangle_m
}{A_0},
\label{eq:delta_power_decomp}
\end{equation}
whose first term is the coherent loss to CP deficit and Doppler and whose second is the ISI excess from symbol-to-symbol power variation. 
Let $\Gamma$ denote the ratio of the focused image floor to its peak, and let $\mathrm{SNR}_m=|\alpha_m|^2P_m/(N\sigma^2)$ be the per-symbol echo-to-noise ratio. 
Then
\begin{align}
\Gamma_{\mathrm{MF}}
&\simeq
\frac{1}{NM}
\left[
\mu_4-1
+\bar{\Delta}_P
+\frac{\left\langle\mathrm{SNR}_m^{-1}\right\rangle_m}{A_0}
\right],
\label{eq:image_mf}\\
\Gamma_{\mathrm{RF}}
&\simeq
\frac{\nu}{NM}
\left[
\bar{\Delta}_P
+\frac{\left\langle\mathrm{SNR}_m^{-1}\right\rangle_m}{A_0}
\right].
\label{eq:image_rf}
\end{align}
The factor $NM\simeq BT_a/(1+T_{cp}\Delta f)$ is the coherent processing gain accumulated over the
range and azimuth dimensions, and it carries the same CP overhead as the communication rate.

\section{Performance Trade-offs between SAR Imaging and Communication}
\label{sec:tradeoffs}

\subsection{Effective Noise-Equivalent Sigma Zero (ENESZ)}
\label{sec:effnesz}

Conventional SAR systems characterize imaging sensitivity by the noise-equivalent sigma zero (NESZ), the backscatter coefficient at which the target return equals the thermal-noise floor of the image~\cite{martone2014asq,kang2018efficient}. 
The NESZ is built from system parameters such as transmit power, antenna gain, imaging geometry, and receiver noise entering through the radar equation.
Therefore, conventional NESZ is scene independent, although it may vary with the acquisition geometry and position within the swath.

In the data-bearing CP-OFDM SAR imaging, the sidelobes, the ICI, the ISI, and the thermal noise are mutually incoherent and contribute to a distributed floor in the focused image. 
The target response keeps its shape and its resolution, so the image quality is set by the ratio of the coherently retained response to this floor. 
Accordingly, we introduce the ENESZ as an ISAC-SAR imaging quality metric that expresses the total image floor on a backscatter-equivalent scale relative to the coherently retained response.
Unlike the conventional NESZ, the ENESZ can depend on the scene backscatter because the waveform-induced floor is generated by the scene returns themselves.

For the distributed scene, the complex scattering coefficients are modeled as mutually uncorrelated random variables with $\E[|\alpha^{(q)}|^2]$ determined by the local backscatter coefficient.
Conditioned on a fixed realization of the scene, the randomness of the communication payload and thermal noise makes the focused image \eqref{eq:bpa} in a random field, which is decomposed as
\begin{equation}
I(\mathbf r)=I_{\mathrm{sig}}(\mathbf r)+I_{\mathrm{fl}}(\mathbf r),
\label{eq:decomp}
\end{equation}
where $I_{\mathrm{sig}}(\mathbf{r})$ is the response accumulated by the intended component of \eqref{eq:percell} and $I_{\mathrm{fl}}(\mathbf r)$ collects zero-mean floor components arising from data-dependent fluctuations, ICI/ISI, and thermal noise.
Their aggregate second-order contribution is characterized by the swath-averaged image floor $F$.

Let $\sigma^0(\mathbf{r})$ denote the backscatter coefficient, the mean radar cross section per unit ground area at $\mathbf{r}$, with swath mean $\bar\sigma^0=\langle\sigma^0(\mathbf{r})\rangle_{\mathrm{swath}}$, and let $S$ and $F$ denote the swath-averaged powers of $I_{\mathrm{sig}}$ and $I_{\mathrm{fl}}$, the latter averaged over the data payload and the noise. 
The image-domain SINR and the corresponding ENESZ are then defined as
\begin{equation}
\mathrm{SINR}_{\mathrm{img}}
=
\frac{S}{F},
~
\sigma_{\mathrm{eff}}
=
\frac{\bar{\sigma}^0}
{\mathrm{SINR}_{\mathrm{img}}}. 
\label{eq:effnesz}
\end{equation}
Both quantities depend on the scene through $\bar{\sigma}^0$, on the waveform through $T_{cp}$ and the constellation moments $(\mu_4,\nu)$, on the receive filter, and on various system parameters.
Hereafter, only the dependence under study is displayed.

Using the coherent-retention factor $A_0$ in \eqref{eq:A0}, the swath-averaged retained signal power is
\begin{equation}
S
=
N M^2 C A_{\mathrm{IRF}}\,
\bar{\sigma}^0
\mathbb E_{\tilde{\tau}}\!\left[A_0\right],
\label{eq:S_swath}
\end{equation}
where $C$ is a constant from the radar equation \cite{cumming2005digital}, and
\begin{equation}
A_{\mathrm{IRF}}
=
\left(\frac{\rho_r}{\sin\theta_i}\right)
\left(\frac{v}{B_a}\right)
\label{eq:A_IRF}
\end{equation}
is the equivalent ground area of a resolution cell.
Applying the same swath-averaging assumption to the scene-dependent floor terms, let $E_{sc} = C\bar{\sigma}^0 W_g vT_a$ denote the aggregate mean channel power. 
Using the per-symbol floor derived in Section~\ref{sec:sar_imaging} and the receiver-dependent moments in Table~\ref{tab:moments}, the floor of the focused-image is
\begin{equation}
    \begin{split}
        F=&
M E_{sc}
\bigg[
\operatorname{Var}[\chi_n]\,
\mathbb E_{\tilde{\tau}}\![A_0]+
\mathbb E\![|\xi_{n,k}|^2]
\mathbb E_{\tilde{\tau}}
\![A_0\overline{\Delta}_P]
\bigg]\\
&+
MN\sigma^2
\mathbb E\![|\xi_{n,k}|^2]
\langle P_m^{-1}\rangle_m .
    \end{split}
\label{eq:F0_image}    
\end{equation}
The two terms inside the brackets represent the receive sidelobes and the ICI/ISI, respectively, while the final term is the thermal-noise floor after SAR imaging.

By substituting \eqref{eq:S_swath} and \eqref{eq:F0_image} into \eqref{eq:effnesz}, the ENESZ is given by 

\begin{equation}
\label{eq:effective_nesz_MF}
\begin{split}
\sigma^{\mathrm{MF}}_{\mathrm{eff}}(&T_{cp})
= \left(1+\frac{T_{cp}}{T}\right)
\bigg[ \underbrace{\frac{T\sigma^2\langle P_m^{-1}\rangle_m}
{T_a C A_{\mathrm{IRF}}\,\mathbb{E}_{\tilde\tau}[A_0]}}_{\mathrm{thermal\text{-}noise~term}} \\
&\quad + \underbrace{\bar\sigma^0 T_w B_a\left(\mu_4-1
+ \frac{\mathbb{E}_{\tilde\tau}\big[A_0\overline{\Delta}_P\big]}
{\mathbb{E}_{\tilde\tau}[A_0]}\right)}_{\mathrm{scene\text{-}dependent~term}} \bigg],
\end{split}
\end{equation}

\begin{equation}
\label{eq:effective_nesz_RF}
\begin{split}
\sigma^{\mathrm{RF}}_{\mathrm{eff}}(T_{cp})
= \nu (1+&\frac{T_{cp}}{T})
\bigg[ \underbrace{\frac{T\sigma^2\langle P_m^{-1}\rangle_m}
{T_a C A_{\mathrm{IRF}}\,\mathbb{E}_{\tilde\tau}[A_0]}}_{\mathrm{thermal\text{-}noise~term}} \\
&\quad + \underbrace{\bar\sigma^0 T_w B_a\,
\frac{\mathbb{E}_{\tilde\tau}\big[A_0\overline{\Delta}_P\big]}
{\mathbb{E}_{\tilde\tau}[A_0]}}_{\mathrm{scene\text{-}dependent~term}} \bigg].
\end{split}
\end{equation}
The first term is independent of the mean scene backscatter. 
The second term scales with $\bar{\sigma}^{0}$ and comprises the data-payload dependent fluctuation, the CP- and Doppler-induced ICI/ISI.

\subsection{Communication Throughput}
\label{sec:communication}

The communication-side analysis in this work focuses on the deterministic throughput loss introduced by the CP duration.
For each communication configuration, the occupied bandwidth $B$, the number of subcarriers $N$, and the average transmit power are held fixed.
Reliable communication is assumed, and performance is quantified by the data-payload throughput.
Accordingly, the communication metric isolates the deterministic throughput overhead of CP framing rather than link-level reliability effects.
Under this abstraction, the CP affects the communication throughput by increasing the OFDM-symbol duration $T_s=T+T_{cp}$ and thereby reducing the number of data-bearing symbols transmitted per unit time.

Let $b_c$ denote the number of data-payload bits conveyed by one data-bearing subcarrier in one OFDM symbol.
Since each OFDM symbol conveys $Nb_c$ data-payload bits and
$M=\lfloor T_a/(T+T_{cp})\rfloor$ symbols are transmitted during one CPI, the communication throughput is
\begin{equation}
R_{\mathrm{com}}(T_{cp})
=
\frac{MNb_c}{T_a}
\simeq
\frac{Nb_c}{T+T_{cp}}
=
\frac{Bb_c}
{1+T_{cp}/T}.
\label{eq:communication_throughput}
\end{equation}
Thus, $T_{cp}$ determines the communication throughput reduction due to CP framing, while $b_c$ sets the level for a given communication configuration.
It is worth noting that $R_{\mathrm{com}}$ represents the nominal data-payload throughput under a reliable communication-link assumption, rather than achievable rate or link-level throughput including SNR or block error rate.

\subsection{Imaging-Communication Trade-off Design via CP Length}

The CP duration enters the ENESZ in three ways.
The overhead factor $1+T_{cp}/T$, which is the same factor that reduces $R_\mathrm{com}$ in \eqref{eq:communication_throughput}, multiplies both terms. 
The thermal-noise term decreases with $T_{cp}$ because $\mathbb{E}_{\tilde\tau}[A_0]$ in its denominator increases. 
The scene-dependent term decreases with $T_{cp}$ because the CP-deficit part of $\bar\Delta_P$ shrinks as the deficit is removed.
Here, the constellation statistics and the normalized Doppler trajectory $\varepsilon(\eta)$ are $T_{cp}$-invariant. 
As $T_{cp}$ increases toward $T_w$, the imaging gains saturate once the swath is fully captured, while the overhead keeps growing.
Therefore, the ENESZ is not always minimized at $T_{cp}=T_w$, and even for imaging alone $T_{cp}$ is a quantity to be optimized rather than set to the swath delay.

In \eqref{eq:effective_nesz_MF}-\eqref{eq:effective_nesz_RF}, the RF removes the pedestal sidelobe but scales the thermal and ICI/ISI floor by $\nu$, while the MF yields the ENESZ including the sidelobes from the scatterers proportional to $\mu_4-1$. 
The floor decreases with $T_{cp}$ and, relative to the pedestal, with the swath mean of backscatter coefficient $\bar\sigma^0$. 
Because the RF scales both the thermal-noise and ICI/ISI terms by $\nu$, the MF tends to be favored when thermal noise or residual CP-deficit interference is dominant.
As $T_{cp}$ increases and the scene-dependent floor becomes increasingly governed by the MF pedestal, the RF can become advantageous by eliminating the $\mu_4-1$ term.
The MF-RF crossover, however, depends jointly on $T_{cp}$, the mean scene backscatter $\bar{\sigma}^0$, and the constellation statistics.

\begin{figure}[t]
\centering
\includegraphics[width=0.85\columnwidth]{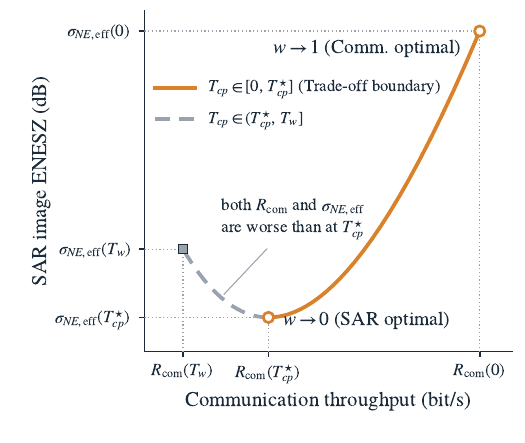}
\caption{Performance trade-off between communication throughput and SAR image ENESZ depending on guard interval (CP length).}
\label{fig:tradeoff}
\end{figure}

The trade-off between communication throughput and SAR image quality can be flexibly controlled by selecting $T_{cp}$ that maximizes their weighted sum,
\begin{equation}
T_{cp}^{\mathrm{opt}}=\arg\max_{T_{cp}}\
w\,\frac{R_{\rm com}(T_{cp})}{R_{\rm com}(0)}
-(1-w)\,\frac{\sigma_{\mathrm{eff}}(T_{cp})}
{\sigma_{\mathrm{eff}}(T_{cp}^*)},
\label{eq:design_w}
\end{equation}
where both metrics are normalized by their respective reference extrema, and $w\in[0,1]$ is a priority weight that sets the priority of communication throughput over image quality. 
Since the objective is not concave in general, \eqref{eq:design_w} is solved by a one-dimensional search over $[0,T_w]$.

Varying the CP duration from $T_{cp}=0$ to $T_{cp}=T_w$ traces a curve in the $(R_{\rm com},\sigma_{\mathrm{eff}})$ plane, providing a geometric representation of the optimization in \eqref{eq:design_w}.
As shown in Fig.~\ref{fig:tradeoff}, the orange branch $0\le T_{cp}\le T_{cp}^{\star}$, where $T_{cp}^{\star}$ minimizes $\sigma^{\mathrm{best}}_{\mathrm{eff}}$ over $[0,T_w]$, forms the trade-off boundary: increasing the CP along this branch lowers the ENESZ at the cost of communication throughput.
Beyond $T_{cp}^{\star}$, both metrics worsen, and the corresponding gray branch is therefore dominated.
The exact trade-off curve and the resulting $T_{cp}^*$ and $T_{cp}^{\mathrm{opt}}$ depend on the system, scene, and communication configurations.

\section{Simulation Results}

\label{sec:model:geometry_signal}
\begin{table}[!t]
\caption{Operating Point (X-band VLEO CP-OFDM SAR system)}
\label{tab:operating_point}
\centering
\begin{tabular}{l l l}
\toprule
Symbol & Quantity & Value \\
\midrule
$h$ & orbital altitude & 250 km \\
$\theta_i$ & incidence angle & $37.78^\circ$ \\
$R_0$ & reference slant range & 316.3 km \\
$v$ & effective velocity & 7611 m/s \\
$f_c$ & carrier frequency & 9.6 GHz \\
$D_a$ & azimuth aperture & 2.4 m \\
$B_a$ & Doppler bandwidth & 5619.5 Hz \\
$T_a$ & CPI & 479.1 ms \\
$\Delta f$ & subcarrier spacing & 30 kHz \\
$N$ & number of subcarriers & 3360 \\
$B$ & bandwidth ($N\Delta f$) & 100.8 MHz \\
$T$ & useful symbol duration ($1/\Delta f$) & 33.33 $\mu$s \\
$T_{cp}$ & CP duration  &  Varied \\
$T_w$& sufficient CP & 28.55 $\mu$s\\
$T_s$ & total symbol period & $T+T_{cp}$ \\
$\theta_{el}$ & elevation beamwidth & $1.00^\circ$ \\
$W_g$ & ground swath & 6.99 km \\
$M$ & symbols per aperture & $\lfloor T_a/T_s\rfloor$ \\
$\rho_r$ & nominal slant range resolution $c/(2B)$ & 1.487 m \\
$\rho_a$ & nominal azimuth resolution $v/B_a$ & 1.354 m \\
$\sigma_{\mathrm NE}$ & noise-equivalent sigma zero (NESZ) & -25.2 dB\\
$\bar{\sigma}^0$ & averaged backscatter-coefficient & -8.14 dB\\
\bottomrule
\end{tabular}
\end{table}

\begin{figure}[t]
\centering
\includegraphics[width=\columnwidth]{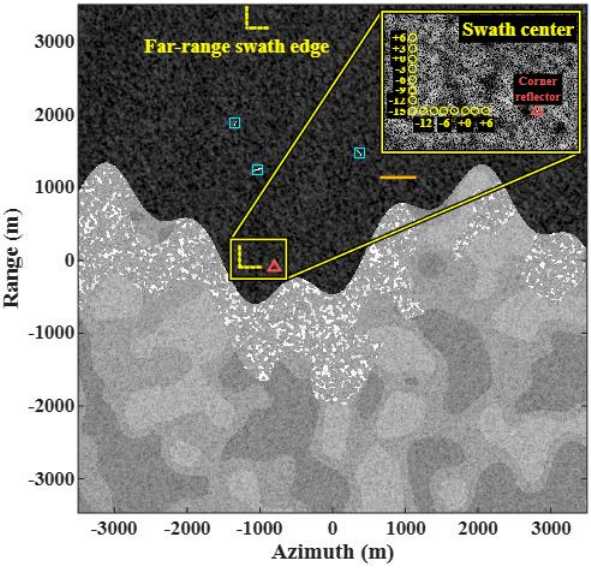}
\caption{Point-scatterer model of the coastal scene with the inset enlarging the boxed area as the RCS per $2~\mathrm{m}\times2~\mathrm{m}$ pixel within $\pm12$~dB of its mean over the water.}
\label{fig:scene}
\end{figure}

\begin{figure}[!t]
\centering
\includegraphics[width=0.85\columnwidth]{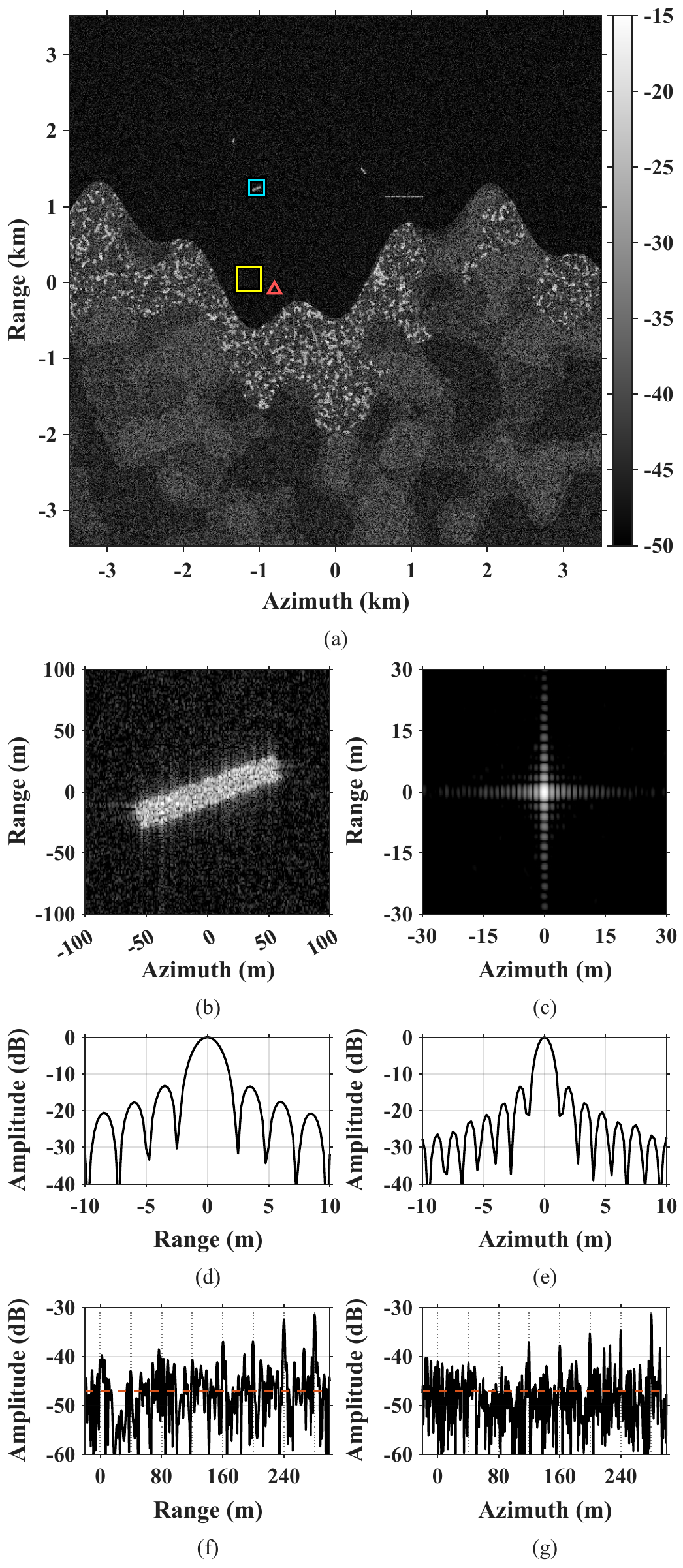}
\caption{SAR image result of the scene in Fig.~\ref{fig:scene} in dB relative to the CR peak for MF, $T_{cp}=T_w$, and 64-QAM, showing (a) the full scene with the ship, the CR, and the swath-center array marked by the cyan box, red triangle, and yellow box, (b) and (c) enlarged views of the ship and the CR, (d) range and (e) azimuth cuts through the CR peak, and (f) range-arm and (g) azimuth-arm cuts of the swath-center array, with dotted lines at the true target positions and a dashed line at the floor.}
\label{fig:imaging result}
\end{figure}

In this section, simulation results are presented to verify the proposed analytical framework and to quantify the trade-off between SAR image quality and communication throughput in the VLEO CP-OFDM ISAC-SAR system. 
The system parameters used throughout the simulations are listed in Table~\ref{tab:operating_point}, which represents an X-band VLEO operating point \cite{vleo2026park}. 
The thermal noise level is specified by the conventional NESZ $\sigma_{\mathrm NE}$, defined at $R_0,~T_{cp}=0,~A_0=1$ in \eqref{eq:effective_nesz_MF}.
It subsumes the radar constant $C$ in \eqref{eq:S_swath}, which represents the system-level link budget.
For these parameters, the swath delay span is shorter than the useful symbol duration, $T_w<T$, so each echo contributes at most one adjacent symbol to the DFT window. 
The CP duration $T_{cp}$ is varied from $0$ to $T_w$, the data symbols are drawn from QPSK, 16-, 64-, and 1024-QAM constellations, and both the MF and the RF receivers are evaluated.

The imaging simulations use a point-scatterer model emulating a coastal area (Fig.~\ref{fig:scene}).
The distributed clutter is represented by one scatterer per resolution cell. 
Its reflectivity is circular Gaussian with mean power proportional to the local backscatter coefficient $\sigma^0$, whose area average over the scene is $\bar\sigma^0$, and stays fixed over the synthetic aperture. 
Land with an urban band along the coast occupies the near range, calm sea occupies the far range, and ship-like bright patches (cyan squares) and a breakwater (orange line) lie on the water. 
Two L-shaped arrays of point targets are placed on the water, one at the swath center and one at the far-range swath edge. 
Along each arm the targets are 40~m apart, and their RCS rises in 3~dB steps from $-15$~dBsm at the corner to $+6$~dBsm, as labeled in the inset. 
This RCS range brackets the level at which the floor masks a point target, so each array shows which targets remain visible above the floor. 
A corner reflector (CR) next to the swath-center array serves as a reference target for verifying the SAR imaging.

A representative imaging result is obtained using the MF with $T_{cp}=T_w$ and 64-QAM, as shown in Fig.~\ref{fig:imaging result}.
With $T_{cp}=T_w$, the CP covers the swath delay span, so the image is free of ISI and its floor consists of the thermal noise and the pedestal in \eqref{eq:effective_nesz_MF}.
The reconstructed scene in Fig.~\ref{fig:imaging result}(a) reproduces the coastal scene of Fig.~\ref{fig:scene}, including the coastline, urban band, ship-like patches, and breakwater at their true positions.
The enlarged views in Fig.~\ref{fig:imaging result}(b) and (c) further show that the ship and the CR are well focused.
The range and azimuth cuts of the CR in Fig.~\ref{fig:imaging result}(d) and (e) have 3~dB widths of 2.17~m in ground range and 1.22~m in azimuth, consistent with $\rho_r/\sin\theta_i$ and $\rho_a$ in Table~\ref{tab:operating_point}.
Moreover, the shape of Fig.~\ref{fig:imaging result}(d) and (e) follows sinc function confirming near-ideal compression in both dimensions.

The floor, in contrast, arises collectively from the thermal noise and the pedestal sidelobes of all scatterers in the scene, and its level determines which targets remain visible.
Along the two arms of the swath-center array in Fig.~\ref{fig:imaging result}(f) and (g), whose RCS rises in 3~dB steps, the local background around the array lies at $-46.9$~dB relative to the CR peak. 
The strongest targets rise well above the floor, the weakest sink into it, and the intermediate targets, whose peaks lie within the fluctuation of the floor, are partially discernible.
Under a CP deficit, however, target visibility is additionally affected by the range-dependent coherent retention, since the capture factor $\beta$ varies with the differential delay across the swath.
The resulting variation of the target peaks relative to the image floor is examined for the swath-center and far-range-edge arrays.

\begin{figure}[t]
\centering
\includegraphics[width=0.85\columnwidth]{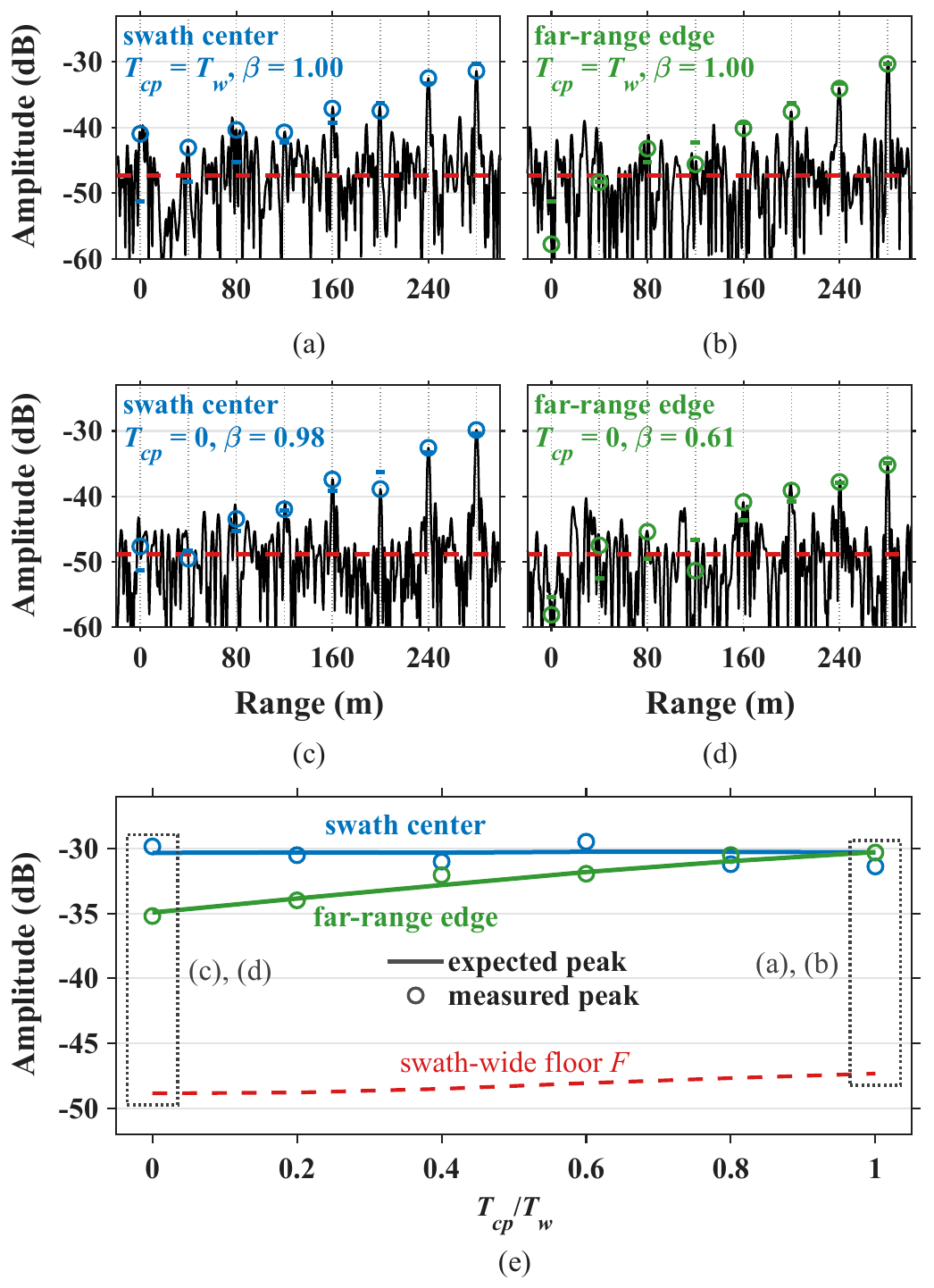}
\caption{Range-dependent effect of CP deficit on the target-array responses.
(a), (b) Range-arm profiles at the swath center and far-range edge for $T_{cp}=T_w$.
(c), (d) Profiles for $T_{cp}=0$, with $\beta=0.98$ and $\beta=0.61$, respectively.
Dotted lines mark the true target positions, circles the measured target peaks, bars the expected peaks, and the red dashed line the swath-wide image floor $F$.
(e) Expected and measured peak amplitudes of the strongest (+6 dBsm) target versus $T_{cp}/T_w$ at the two locations, together with $F$.}
\label{fig:rungs}
\end{figure}

At $T_{cp}=T_w$, the swath-center and far-range-edge arrays are both fully captured, and their measured target peaks closely follow the expected levels as shown in Fig. \ref{fig:rungs}(a) and (b).
When the CP is removed, the capture factor remains close to unity at the swath center $(\beta=0.98)$ but decreases to $0.61$ at the far-range edge (Fig. \ref{fig:rungs}(c) and (d)).
Consequently, the far-range target peaks undergo a pronounced reduction, whereas those near the swath center remain nearly unchanged.
This spatial dependence persists over the full CP range.
The strongest target at the swath center experiences only negligible coherent-retention loss because its capture factor $\beta$ remains close to unity, whereas the corresponding far-range target gains about 5~dB as $T_{cp}$ increases from $0$ to $T_w$ (Fig. \ref{fig:rungs}(e)).
The measured peaks closely follow the expected values over the CP range, validating the range-dependent coherent-retention model.
While this coherent-retention loss varies locally across the swath, the overall imaging performance is governed jointly by the retained signal and the aggregate image floor.
The impact of the second impairment in the unified model, platform-induced Doppler ICI, is considered next.

\begin{figure}[t]
\centering
\includegraphics[width=0.85\columnwidth]{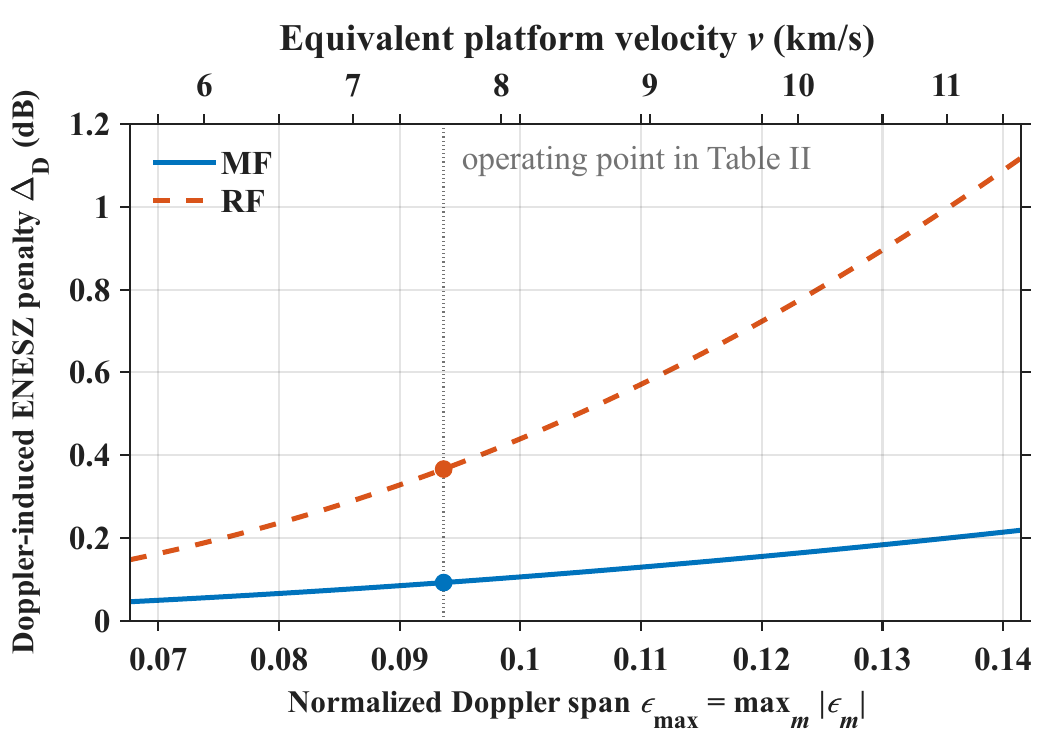}
\caption{Doppler-induced ENESZ penalty as a function of the normalized Doppler span $\varepsilon_{max}=\max_m |\varepsilon_m|$ for the MF and RF with 64-QAM and $T_{cp}=T_w$.
The markers indicate the operating point in Table~\ref{tab:operating_point}.}
\label{fig:vel_sweep}
\end{figure}

The contribution of platform-induced Doppler is further isolated by comparing the ENESZ with that of corresponding zero-Doppler case. 
We define the resulting Doppler-induced penalty as
\begin{equation}
   \Delta_\mathrm D = 10\log_{10} \frac{\sigma_{\rm eff}(\varepsilon_m)} {\sigma_{\rm eff}(\varepsilon_m=0)}, 
\end{equation}
with all other conditions kept unchanged. The penalty increases monotonically with the normalized Doppler span \(\varepsilon_{\max}=\max_m|\varepsilon_m|\), consistent with the coupling-energy redistribution in Fig. \ref{fig:partition}(b).
At the operating point in Table \ref{tab:operating_point}, \(\varepsilon_{\max}\simeq0.094\), corresponding to \(v=7.61\) km/s, the penalty is approximately \(0.1\) dB for the MF and \(0.37\) dB for the RF (Fig. \ref{fig:vel_sweep}). 
These results confirm that platform-induced ICI introduces a velocity-dependent ENESZ penalty that increases with the normalized Doppler span.

\begin{figure}[t]
\centering
\includegraphics[width=0.85\columnwidth]{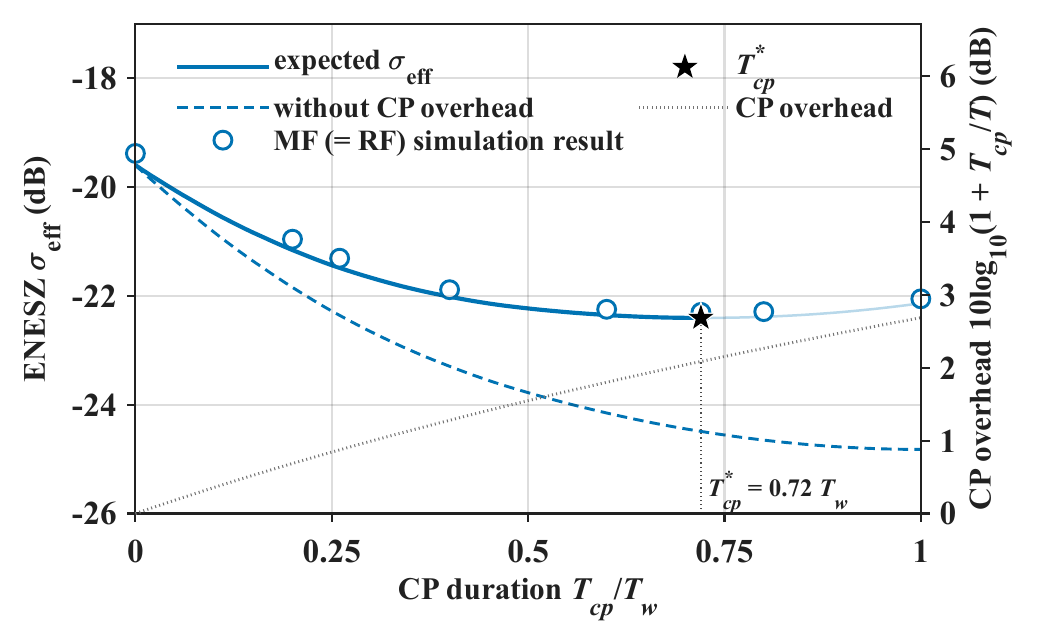}
\caption{CP-dependent ENESZ for QPSK as a function of the normalized CP duration $T_{cp}/T_w$.  
The solid and dashed curves denote the analytical ENESZ with and without the CP-overhead factor $1+T_{cp}/T$, respectively, and the circles denote the simulation results. 
The dotted curve on the right axis gives the CP-overhead penalty $10\log_{10}(1+T_{cp}/T)$, and the star marks the imaging-optimal $T_{cp}^{\star}$.}
\label{fig:cp_qpsk}
\end{figure}

With the range-dependent CP-deficit effect and the platform-induced Doppler penalty characterized, the role of the CP duration itself is first examined using QPSK.
Its constant-modulus property gives $\mu_4=\nu=1$, so that the MF and RF become equivalent and the CP-dependent behavior can be isolated from receiver- and constellation-dependent effects.
As $T_{cp}$ increases from zero, the analytical ENESZ initially decreases because the CP deficit is progressively alleviated, increasing the coherent-retention factor while reducing the associated ICI/ISI contribution.
The simulation results displayed in Fig.~\ref{fig:cp_qpsk} closely follow the analytical prediction over the entire CP range.

The competing role of the CP overhead can be seen by removing the multiplicative factor $1+T_{cp}/T$ from \eqref{eq:effective_nesz_MF}.
The resulting ENESZ decreases monotonically with $T_{cp}$, showing that the intrinsic imaging benefit continues as the CP deficit is reduced.
In contrast, the CP-overhead penalty $10\log_{10}(1+T_{cp}/T)$ increases monotonically because a longer symbol period reduces the number of OFDM symbols available for coherent integration.
The balance between these two effects produces an interior minimum at
$T_{cp}^{\star}=0.72T_w$, rather than at the sufficient-CP value $T_w$.
Thus, even when the sensing waveform is free of constellation-dependent pedestal sidelobes, fully covering the swath delay does not minimize the SAR image ENESZ for the considered configuration.

\begin{figure}[t]
\centering
\includegraphics[width=0.85\columnwidth]{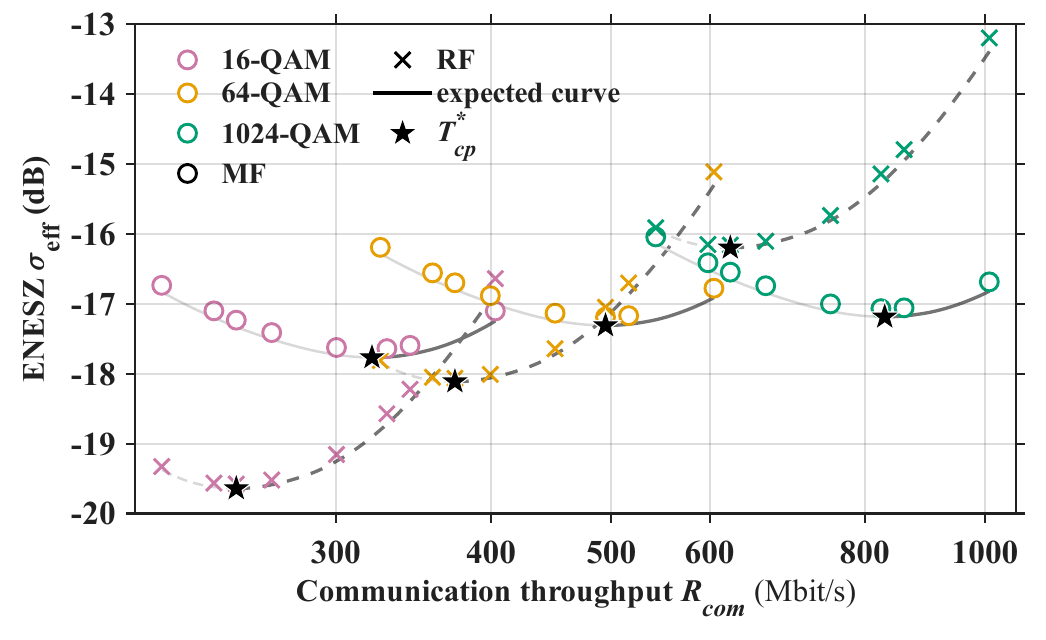}
\caption{SAR imaging-communication throughput trade-off obtained by varying $T_{cp}$ from $0$ to $T_w$ for 16-, 64-, and 1024-QAM. 
Circles and crosses denote the MF and RF cases, respectively, and the curves show the analytical predictions. 
Stars indicate the SAR-optimal $T_{cp}^{\star}$ independently obtained for each constellation and receive filter.}
\label{fig:ENESZ}
\end{figure}

Higher-order QAM introduces both constellation-dependent imaging effects and increased communication throughput.
The resulting imaging-communication trade-off is examined by varying $T_{cp}$ over $0\leq T_{cp}\leq T_w$ for 16-QAM, 64-QAM, and 1024-QAM, while all other system parameters are fixed to those in Table~\ref{tab:operating_point}.
For each constellation and receive filter, the rightmost operating point corresponds to $T_{cp}=0$.
As $T_{cp}$ increases, the operating point moves leftward because the communication throughput decreases monotonically with the CP overhead, whereas the ENESZ initially improves as the CP deficit is alleviated.
The simulated operating points closely agree with the analytical predictions of \eqref{eq:effective_nesz_MF}-\eqref{eq:communication_throughput}, validating the proposed ENESZ formulation after the end-to-end SAR imaging process (Fig.~\ref{fig:ENESZ}).

For the RF, all QAM constellations used in this study attain their minimum ENESZ at the same normalized CP duration, $T_{cp}^{\star}=0.72T_w$, identical to the QPSK optimum in Fig.~\ref{fig:cp_qpsk}.
This invariance follows directly from \eqref{eq:effective_nesz_RF}, where the constellation-dependent factor $\nu$ multiplies the entire ENESZ independently of $T_{cp}$.
The constellation therefore changes the RF ENESZ level but not the location of its minimum.

In contrast, the MF optimum depends on the constellation through the pedestal sidelobe $\mu_4-1$ in \eqref{eq:effective_nesz_MF}.
The resulting $T_{cp}^{\star}$ decreases from $0.30T_w$ for 16-QAM to $0.26T_w$ for 64-QAM and $0.25T_w$ for 1024-QAM.
As the pedestal becomes larger, a greater fraction of the image floor becomes irreducible by increasing the CP, while the CP overhead continues to reduce the coherent processing gain.
Consequently, the benefit of lengthening the CP saturates earlier, shifting the imaging-optimal operating point toward a shorter CP.
This behavior differs from the RF case, in which the constellation changes only the ENESZ scale and not the normalized CP duration at the optimum.

Overall, the numerical results validate the proposed analytical framework from the local target response to the swath-averaged ENESZ.
The CP deficit produces a range-dependent coherent-retention loss, whereas platform-induced Doppler ICI introduces a velocity-dependent penalty that grows with the normalized Doppler span.
Together with the CP overhead, these effects lead to a nontrivial imaging-communication trade-off whose optimum depends on the receive filter and the communication constellation.

\section{Conclusion}

This paper developed an end-to-end analytical framework for data-bearing CP-OFDM ISAC-SAR in a VLEO environment.
A unified received-signal model jointly accounts for random communication payloads, range-dependent CP deficit, and platform-induced Doppler, and propagates their effects through receive filtering, range compression, and azimuth focusing.
The resulting image-domain analysis separates the coherently retained response from data-dependent sidelobes, ICI/ISI, and thermal noise, enabling the formulation of an ENESZ that places their combined impact on a common backscatter-equivalent scale.

The analysis shows that CP deficit and platform-induced Doppler affect the focused image through distinct but coupled mechanisms: the former produces range-dependent coherent-retention loss and ICI/ISI across the swath, whereas the latter introduces slow-time-dependent ICI over the synthetic aperture.
The CP duration controls not only the suppression of delay-induced interference but also the coherent processing gain and communication throughput.
As a result, the sufficient-CP duration does not, in general, coincide with the imaging-optimal CP duration, which further depends on the receive filter and communication constellation.
End-to-end simulations validate the derived image-domain statistics and ENESZ predictions and confirm the resulting CP-dependent imaging-communication trade-off.
These results provide an analytical basis for system design in data-bearing CP-OFDM VLEO ISAC-SAR that jointly accounts for waveform, platform, and receiver effects.

\bibliographystyle{IEEEtran}
\bibliography{taes_refs_1}

\end{document}